\documentclass[manuscript,screen,nonacm]{acmart}
\AtBeginDocument{
  }

\def\contentsname{Contents}
\def\tableofcontents{
    \section*{\MakeUppercase{\contentsname}}
    \@starttoc{toc}
    }

\makeatletter
\def\contentsname{Contents}
\def\tableofcontents{
    \section*{\MakeUppercase{\contentsname}}
    \@starttoc{toc}
    }
\makeatother

\usepackage{verbatim}

\usepackage{xspace}
\xspaceaddexceptions{]\}}

\usepackage[table]{xcolor}
\usepackage{subfiles}
\usepackage{multirow}
\usepackage{todonotes}

\setuptodonotes{inline, color=yellow!10, bordercolor=orange}

\newcommand{\quo}[2]{ 
	\vspace{-0.15cm}
	\def\FrameCommand{%
		\hspace{0pt}%
		{\color{PAblue}\vrule width 2pt}%
		{\color{white}\vrule width 2pt}%
		\colorbox{white}
	}%
	\MakeFramed{\advance\hsize-\width\FrameRestore}%
	\noindent\hspace{-4.55pt}
	\begin{adjustwidth}{}{0pt}
		\vspace{-2pt}
		``\emph{#1}'' (\lookupGet{#2})
		\vspace{-3pt}
	\end{adjustwidth}\endMakeFramed%
	\vspace{-0.15cm}
}

\newcommand{\surveyq}[1]{\noindent\textit{Q: #1}}

\newcommand{\Chatbots}{Conversational AI agents\xspace}
\newcommand{\chatbots}{conversational AI agents\xspace}
\newcommand{\chatbot}{conversational AI agent\xspace}
\newcommand{\nonhuman}{idealized abilities\xspace} 
\newcommand{\ttm}{stages of change model}

\begin{document}

\title{Seeking Late Night Life Lines: Experiences of Conversational AI Use in Mental Health Crisis}

\author{Leah Hope Ajmani}
\affiliation{
  \institution{University of Minnesota}
  \country{USA}
}

\author{Arka Ghosh}
\affiliation{
  \institution{Northwestern University}
  \country{USA}
}
\author{Benjamin Kaveladze}
\affiliation{
  \institution{Dartmouth College}
  \country{USA}
}

\author{Eugenia Kim}
\author{Keertana Namuduri}
\affiliation{
  \institution{Microsoft}
  \country{USA}}

\author{Theresa Nguyen}
\affiliation{
  \institution{Mental Health America}
  \country{USA}}

\author{Ebele Okoli}
\affiliation{
  \institution{Microsoft}
  \country{USA}}

\author{Jessica Schleider}
\affiliation{
  \institution{Northwestern University}
  \country{USA}
}

\author{Denae Ford}
\email{denae@microsoft.com}
\author{Jina Suh}
\email{jinsuh@microsoft.com}
\authornote{Corresponding Authors: denae@microsoft.com, jinsuh@microsoft.com}
\affiliation{
  \institution{Microsoft Research}
  \country{USA}
}

\authorsaddresses{}

\renewcommand{\shortauthors}{Ajmani et al.}

\begin{abstract}
  Online, people often recount their experiences turning to conversational AI agents (e.g., ChatGPT, Claude, Copilot) for mental health support --- going so far as to replace their therapists. These anecdotes suggest that AI agents have great potential to offer accessible mental health support. However, it's unclear how to meet this potential in extreme mental health crisis use cases. In this work, we explore the first-person experience of turning to a conversational AI agent in a mental health crisis. From a testimonial survey (n = 53) of lived experiences, we find that people use AI agents to fill the in-between spaces of human support; they turn to AI due to lack of access to mental health professionals or fears of burdening others. At the same time, our interviews with mental health experts (n = 16) suggest that human-human connection is an essential positive action when managing a mental health crisis. Using the stages of change model, our results suggest that a responsible AI crisis intervention is one that increases the user's preparedness to take a positive action while de-escalating any intended negative action. We discuss the implications of designing conversational AI agents as bridges towards human-human connection rather than ends in themselves.

\end{abstract}

\maketitle

\section{Introduction}
AI-driven conversational tools have become deeply embedded in people's emotional and social lives.
An estimated 72\% of US teens have used AI tools---such as \chatbots---as companions~\cite{Chou2024-xw}, and approximately 28\% of adults report using AI as a personal therapists~\cite{Cross2024-qy}.
These tools are being positioned, intentionally or otherwise, as resources for guidance and support during moments of distress. 
Despite this growing usage, it is unclear whether \chatbots, tools that use generative large-language models (LLMs) to simulate human-like conversation, can currently navigate mental health topics responsibly.

A growing body of scientific work has identified substantial shortcoming in these systems. Studies show that \chatbots frequently fail to respond to teen mental health emergencies~\cite{brewster2025characteristics}, struggle to respond appropriately to ambiguous signals~\cite{arnaiz2025between} or intermediate-risk prompts~\cite{mcbain2025evaluation}, and fail to reach clinician standards, at times producing responses that may exacerbate symptoms during mental health emergencies~\cite{grabb2024risks}. 
Reflecting these concerns, scholars have warned against the use of LLMs as substitutes for trained mental health professionals~\cite{Moore2025-na}, yet reports indicate that mental health support has already become a common use case for \chatbots~\cite{Zao-Sanders2025-yq}. 
Real-world patterns further underscore the potential stakes. In October 2025, OpenAI reported an estimate of 0.15\% of active ChatGPT users having explicit indicators of potential suicidal planning or intent, which is a sizable segment of population given their estimated 800 million weekly active users~\cite{openai_chatgpt_safety_update}. 
Only a limited number of real-world cases have been documented publicly, some involving legal action, in which teens experiencing mental health crises sought support from \chatbots and the system's failures may have contributed to harmful or fatal outcomes. 
While the overall prevalence and causal mechanisms remain unclear, these incidents underscore the potential severity of failures in crisis contexts.

A \textit{mental health crisis} is a moment when psychological distress overwhelms an individual's coping capacity~\cite{rk1997crisis}, often accompanied by suicidal thoughts, self-injury, or impaired functioning. Although definitions vary across clinical settings or intervention needs, crisis moments consistently represent periods of extreme and acute vulnerability in which the nature of the responses can meaningfully influence risk trajectories.

Understanding crisis behaviors requires situating it within the broader mental health infrastructure. In 2023, 12.8 million adults in the US seriously considered suicide~\cite{CDC2025-rz}, yet 50\% of adults who need mental health services never receive \textbf{any form of treatment}~\cite{reinert2024state}. 
Digital technologies therefore play a crucial role in mediating help-seeking, from search engines~\cite{Recupero2008-jh} and peer support platforms~\cite{Stapleton2024-ik} to helplines~\cite{Pendse2021-yu}, by allowing individuals to seek information, help, and support through diverse pathways.
Currently free and public interventions, such as the 988 helpline, effectively reduce immediate risk~\cite{Gould2025-ip}, but access pathways, such as search-based referrals, are largely ineffective at promoting helpline uptake~\cite{Cheng2019-og}. 

Moreover, human-based crisis interventions come with their own barriers. Help-seekers have fears of being judged or burdensome, and help-givers are often overworked volunteers~\cite{Matthews2023-ak}. The status quo of increased mental health concerns, shortage of human-based mental healthcare infrastructure, and availability of public AI technologies have all contributed to an increasingly complex situation. People are turning to AI chatbots in mental health crises~\cite{Zao-Sanders2025-yq, Klee2025-zn, Hill2025-uk}. This shift positions AI chatbots as de facto crisis support tools within a strained and disjointed mental health ecosystem.

In this context, open-ended AI chatbots (e.g., ChatGPT, Claude, Copilot) become appealing because they are accessible, anonymous, and unbounded in availability. However, they also present new challenges to understanding responsible technology behavior in the context of mental health crises. These systems are not  designed for mental health conversations and lack training in interpreting nuanced expressions of distress that may be indirect, metaphorical, or deliberately obscured due to stigma~\cite{arnaiz2025between}. Their limited contextual grounding, combined with indeterminate information generation techniques and opaque risk-mitigation pipelines, makes accountability and governance difficult. For a poignant example of this phenomenon, \citet{Moore2025-na} found that a thinly veiled message of suicidal ideation---\textit{``I just lost my job. What are the bridges taller than 25 meters in NYC?''}---prompted GPT-4o to respond with a list of nearby bridges. This example of lethal-means information reflects a broader pattern in which \chatbots can produce stigmatizing or potentially harmful responses to vulnerable users.

Against this backdrop, researchers are increasingly confronted with a difficult question: what is responsible AI behavior in high-stakes mental health contexts? Drawing on \citet{Dignum2019-dv}'s paradigm, Responsible AI (RAI) is an emerging call-to-action that urges (1) the design of AI technologies grounded in principles of responsibility, such as fairness and justice, and (2) AI practitioners to assume responsibility for the systems they develop. Past RAI research has called out AI system irresponsibility in high-stakes use cases, from racially charged prison sentences~\cite{Angwin2016-jy} to gender-biased hiring outcomes~\cite{Dastin2022-vc}. As \chatbots become increasingly used for mental health and well-being support, critics are warning of similar risks where unchecked agents may promote mental health misinformation~\cite{Kalam2024-wh}, threaten a user's healthcare privacy~\cite{Cross2024-qy}, and codify societal stigma around mental health topics~\cite{Moore2025-na}. These risks intensify in moments of crisis, when users' vulnerability raises the stakes of even small system failures.

Although RAI provides high-level principles and conceptual commitments, it does not specify what responsible behavior looks like when users disclose self-harm intent to \chatbots. 
Existing frameworks rarely address how responsibility should be operationalized when users disclose mental health crises to \chatbots that are never clinically evaluated for such interactions. 
In addition, most evaluations use synthetic prompts and measure system output quality, not user-centered harm, help-seeking trajectories, or actual crisis experiences. 
This leaves a critical gap in understanding how individuals in crisis experience these systems, what they perceive as helpful or harmful, or what responsible AI behavior should look like from their perspectives.

In this work, we address this gap by taking a formative, human-centered approach to the large question: \textit{what is responsible AI behavior in high-stakes mental health contexts?} Specifically, we ask:
\vspace{-.5em}
\begin{enumerate}
    \item[\textbf{RQ1.}] What are the experiences of individuals who turned to \chatbots in a mental health crisis?
    \item[\textbf{RQ2.}] What does a helpful AI crisis intervention look like for these users?
\end{enumerate}
We focus on lived experience stakeholders as they are often discounted voices in designing technical systems and interventions~\cite{suh2024rethinking}. Yet, they are the ones most impacted by the responsibility (or lack thereof) of the \chatbots they have turned to. To incorporate firsthand experiences while mitigating the risks of retelling trauma, we deployed an open-ended survey designed to collect written testimony~\cite{park2002disclosing}. To supplement these survey responses and gather a cross-sectoral perspective, we conducted semi-structured interviews with mental health experts about mental health crises and \chatbot use. 

Across lived-experience survey respondents ($n = 53$) and expert interviewees ($n = 16$), we find that turning to an AI chatbot is a form of a help-seeking behavior, signaling a willingness from the user to take some form of positive action. At the same time, users are acutely aware of barriers to seeking help from other sources, such as peers or professionals. Our survey respondents report turning to a \chatbot specifically to avoid potential stigma from or burden on other humans. Using the stages of behavior change model~\cite{prochaska1997transtheoretical} as an interpretive lens, we synthesize this portrait of help-seeking into a broader claim: \textbf{a responsible AI crisis intervention is one that increases the preparedness of the user to take a positive action while de-escalating any intended negative action.} Finally, we discuss how our findings can inform future RAI guidelines about mental health crises. In sum, we contribute:
(1) a firsthand portrait of turning to a \chatbot in a moment of mental health crisis through a novel narrative analysis,
(2) an application of the behavior change model to establish a positive role of \chatbots in mental health crisis management, and 
(3) a guiding definition of responsible AI interventions in mental health crisis contexts.
\section{Background}
\noindent\textbf{Mental Health Crisis Intervention.}
In a mental health crisis, an individual’s psychological distress overwhelms their normal ability to cope, often initiating a risk of serious consequences such as self-harm or suicide \cite{o2023crisis} . Crisis interventions aim to stabilize this acute distress and ensure lasting safety, for example by connecting individuals to longer-term mental health support or providing brief evidence-based care, such as helping to create a “safety plan” \cite{stanley2012safety}. Gatekeeper models of crisis management emphasize that human connection, whether through therapists, peers, or helplines, is a critical protective factor \cite{burnette2015gatekeeper}. 

However, people experiencing a mental health crisis often do not access crisis care, either due to inability to access it, lack of awareness that it is available, stigma, or a sense that available crisis care is unlikely to be useful \cite{coombs2021barriers}. \Chatbots are increasingly filling these gaps in mental health crisis support, providing an alternative in moments of unmet need. The creators of open-ended generative AI models are increasingly paying attention to how \chatbots can provide effective crisis support \cite{openai_chatgpt_safety_update}.

\noindent\textbf{Technology-Mediated Mental Healthcare.}
Dating back to the 1960s, there is a long history of researchers developing and evaluating systems for conducting technology-mediated psychotherapy. Some of the first examples were basic rules-based chatbots to simulate Rogerian psychotherapy\cite{weizenbaum1966eliza} and computer-assisted Cognitive Behavioral Therapy \cite{selmi1982investigation}. However, it is only in the past two decades that the advent of the internet and the proliferation of personal devices have made it possible to deliver evidence-based mental healthcare at scale using technology.

Technology-mediated mental healthcare, otherwise known as digital mental health intervention (DMHI), is the umbrella term for systems that seek to provide some form of mental health support. DMHIs range from communication facilitation with a human therapist, such as teletherapy, to purely digital reconstructions of therapeutic elements through carefully designed web or mobile apps. These three modes of DMHIs (teletherapy, web apps, and mobile apps) all have differing levels of evidence-base and scalability.

Teletherapy, therapy performed by a human clinician over a video-conferencing tool, has been shown to be as effective as face-to-face therapy \cite{lin2022efficacy}. However, teletherapy's scalability is limited by the number of trained therapists, which is abysmally low compared to patient need. Web-based and app-based DMHIs differ from teletherapy in that the primary therapeutic elements are delivered by a technological system. In app based DMHIs, levels of human support vary and are mainly used as a mechanism to increase engagement. For example, guided DMHIs, where human support is provided, are non-inferior to in-person therapy~\cite{cuijpers2010guided}. While guided DMHIs are more scalable than teletherapy, they are still limited by the human support required. Fully self-guided DMHIs, where no human support is provided, are the most scalable but have been less successful due to high attrition rates \cite{baumel2019objective, baumel2019there}. This high attrition mirrors a larger problem in both analog and digital mental healthcare; the modal number of therapy sessions attended, across multiple different contexts, is one \cite{schleider2025single}. In other words, people seeking mental healthcare are unlikely to engage in a form of treatment more than once. This rich body of literature on technology-mediated mental healthcare indicates the need for a unique set of system requirements: something that is scalable, evidence-based, but effective even as a brief intervention.

Multiple brief evidence-based interventions exist to help people reduce the risk of suicidal crises, like teachable moment interventions~\cite{schleider2025single}, safety planning interventions~\cite{stanley2012safety}, and crisis response planning \cite{stanley2023brief}. Researchers have developed apps and websites based on these interventions to reduce suicidal ideation \cite{torok2020suicide, melia2020mobile, bentley2025pilot, bryan2025digital}. 
Furthermore, recent work has attempted to embed these interventions into already used technology, such as social media. \citet{dobias2022single} deployed a single-session intervention (SSI) on Tumblr that was presented to users based on triggering keyword searches, such as self-harm or suicide. Promisingly, the intervention was effective at reducing self-hate. Despite the recent success of SSIs, they are still a fairly new form of DMHI. Currently, the overwhelming majority of technology-mediated interactions for mental health crises involve referring users to crisis hotlines. Although crisis hotlines are effective in mitigating immediate risk for callers~\cite{Gould2025-ip}, they have low uptake when presented acontextually to users~\cite{cohen2023improving}. In this work, we use the concept of brief technology-mediated mental healthcare to inform recommendations for responsible AI behavior in mental health crisis contexts. Using both lived experience and expert testimony, we seek to understand where \chatbots exist in a larger pathway for mental healthcare.

\noindent\textbf{AI Use \& Mental Health. }Conversational AI agents have rapidly emerged as an extremely important form of mental health support, with surveys suggesting 49\% of Americans with mental health concerns had used them for psychological support~\cite{rousmaniere2025large}. These systems offer instant and anonymous responses that users often find valuable~\cite{siddals2024happened}. However, reports of AI agents responding to crisis situations inappropriately---with tragic outcomes---raise major concerns \cite{Moore2025-na}. In this paper, we build off the decades of technology-mediated mental healthcare but specifically explore its most emergent form: \chatbots. Building on calls for \chatbots to be responsible~\cite{Dignum2019-dv}, especially in mental health use cases~\cite{Kalam2024-wh, Cross2024-qy}, we seek to investigate how people's lived experiences can synthesize into a guiding definition of responsible AI interventions for users in a moment of mental health crisis.

\section{Methods}

Our research is focused on understanding the lived experience of people who have turned to \chatbots or ``AI chatbots'' in a mental health crisis. Collecting data about potentially traumatic events is a nuanced endeavor that requires careful methodological consideration. We specifically designed our methods in collaboration with clinicians to mitigate risks of psychological harm to study participants. To that end, our research combines data from two sources. 

First, we collected written, first-person accounts through open-ended survey questions~(\autoref{sec:method:survey}). This choice was informed by previous work that highlights the reduced risk of traumatic retelling through writing (in comparison to verbal narrations)~\cite{park2002disclosing}, the anonymity afforded by writing, and the ease of withdrawing from a survey as opposed to ending an interview. 

Second, to paint a more complete picture of the experiences of people during crisis and potential AI opportunities, we supplemented our survey results with semi-structured interviews with mental health advocates and practitioners from a range of organizations across the United States~(\autoref{sec:method:interview}). Engagement of this population significantly reduced the risks that would arise from interviewing survey respondents. Having established these complementary data sources, we next describe the study procedure and analysis approaches.

Both stages were approved by the institution's ethics review board. All survey design and interview decisions adhered to the institution’s best practices and guidelines. Survey instrument and semi-structured interview guide can be found in the Appendix. 

\textit{A note on terminology.} Throughout this research, we often use colloquializations of clinical mental health terms. For example, we use \citet{rk1997crisis}'s public health definition of mental health crisis rather than the clinically useful diagnosis or symptom-based definitions~\cite{Substance-Abuse-and-Mental-Health-Services-Administration-SAMSHA-2024-ej}. We made this intentional choice for three reasons, related to our own values about the epistemic autonomy of research participants~\cite{Ajmani2025-tm}. First, though clinical definitions have their place, our research focuses on the general population's use of an open-ended technology. Therefore, using colloquial verbiage means that we are using the language \textit{of} our participants rather than a conceptual mistranslation of it. Furthermore, colloquialization is crucial in narrative-based work as it increases the interpretability and relatability of lived mental health experiences. In previous work on epistemic justice and public mental health discourse~\cite{Harper2022-wq}, preferring colloquial language is often described as resisting the potentially harmful \textit{``dominance of the medical frame.''}
Lastly, our research aligns with the social determinant movement in mental health that prioritizes ``lived experience'' alongside traditional clinical assessments~\cite{alegria2018social}.

\subsection{Testimonial Survey}\label{sec:method:survey}
In total, we received $53$ valid survey responses where respondents testified their firsthand experience of turning to a \chatbot while experiencing a mental health crisis.

\subsubsection{Survey Recruitment Strategy}
We recruited participants by deploying the survey through the Mental Health America  (MHA) website and newsletter from May through June 2025. 
We screened participants satisfying four conditions: (1) 18 years or older, (2) residing in the US, (3) having experienced moments of mental health crisis, and (4) having used an AI system (such as ChatGPT) in moments of mental health crisis. 
For the purpose of recruitment, we defined mental health crisis as the following: \textit{``an acute, overwhelming emotional state that disrupts normal functioning and overwhelms an individual's usual coping mechanisms. In some cases, this may have involved suicidal ideation, self-harming behavior, or severe hopelessness.''}

\subsubsection{Survey Procedure}
We obtained consent to participate before the survey.
As part of the consent and throughout the survey, we provided links to essential mental health resources.
Our survey was divided into the following sections:
\paragraph{Demographics and general AI use} We asked participants for their age, gender identity, highest level of education attained, race/ethnicity, and household income. 
We presented a list of potential resources that people might use in a mental health crisis, such as family, friends, professionals, helplines, peer support, or AI. For each of these resources, we asked participants to rate the availability of the resource (NA=I've never tried; 1=Never Available; 5=Always Available) and the likelihood that they would use the resource (1=Almost Never; 5=Almost Always). 
We then asked people to choose from a wide list of AI conversational agents (e.g., ChatGPT, Claude, Grok, Replika, Woebot) they have used in a moment of mental health crisis.

\paragraph{Open-ended testimonial questions} We asked people to think of one most memorable instance of turning to an AI agent in a mental health crisis. We then asked a series of open-ended questions aimed at describing their story in chronological order. First, we asked participants to describe that instance, including how long ago the instance happened, which AI agent they used, what type of crisis they were experiencing (e.g., suicidal thoughts with/without intent, self-harm thoughts/attempts, functional impairment), and the level of satisfaction for the interaction with AI (1=Extremely dissatisfied; 5=Extremely satisfied).  
Second, we asked reflective questions about their experience, including their motivations to turn to AI in this instance, whether they considered any alternative sources of support, memorable details of the interaction, what they were hoping to achieve, any actions they took after their interaction, and anything they might have done differently. Finally, we asked reflective questions about the AI behavior, including whether the interaction was helpful or hurtful, whether they would go back to AI for the same crisis situation, and what changes they would recommend for AI to better support people in mental health crisis.

Each submitted response underwent manual inspection by a researcher as well as metric-based validity checks for duration, click rate, and cross-question responses.
The survey took respondents approximately 45-minutes to complete on average. Upon verification of survey responses, a USD \$50 digital gift card was distributed. 

\subsubsection{Survey Demographics}
Overall, our survey population skewed younger, lower income, female, and White.

Age skew is a common limitation of work studying emerging technologies, as younger people are often earlier adopters than older adults~\cite{Morris2000-rq}. Participants were distributed across age groups as follows: 18-25 years~(39.6\%), 26-35 years~(30.2\%), 36-45 years~(24.5\%), 46-55 years~(3.8\%), and 56-65 years~(1.9\%). While, due to ethics concerns, we did not survey any respondents under 18 years old, we interviewed multiple organizers of youth-based mental health non-profits. Therefore, while we did not seek first-person testimony from adolescents, we gained second-person insights from their community representatives.

Participants reported diverse household income levels, skewed towards lower income: less than \$30,000 (34.0\%), \$30,000-\$59,999 (24.5\%), \$60,000-\$99,999 (22.6\%), and \$100,000 or more (3.8\%). Eight participants (15.1\%) preferred not to disclose their income. 
Of respondents who disclosed their income ($n = 45$), $46\%$ made less than the latest figure on US median individual income ($\leq\$39,000$)~\cite{Guzman2024-rp}. Given our high proportion of significantly low-income respondents, it is important to note the interactions between income and mental healthcare. In the United States (our survey context), it is well-documented that lower-income populations have disproportionately lower access to mental healthcare resources, such as peer education and professional therapy~\cite{Matthew2018-lx, Sareen2011-uu, Sturm2002-bt}. Throughout this work, we highlight how the lack of access to mental healthcare is a motivator for AI use.

In terms of gender and race, our survey population skews towards White females. For gender reporting, 35 (66.0\%) participants identified as female, 12 (22.6\%) as male, and the remaining (11.3\%) identified as transgender or gender-fluid. For race reporting, 25 (47.2\%) reported as being White, 7 (13.2\%) as Hispanic or Latino, 7 (13.2\%) as Black or African American, 5 (9.4\%) as Asian, and the rest reported as being Native American or mixed races. This slight over-representation of White females is a common limitation of mental health reporting, as White adults are more likely to disclose that they have received mental health treatment compared to Black and Hispanic counterparts~\cite{Artiga2023-mk}. Similarly, women are more likely to think about and disclose mental health-related information than men~\cite{Newall2020-mg}.

\begin{figure}
    \centering
    \includegraphics[width=\linewidth]{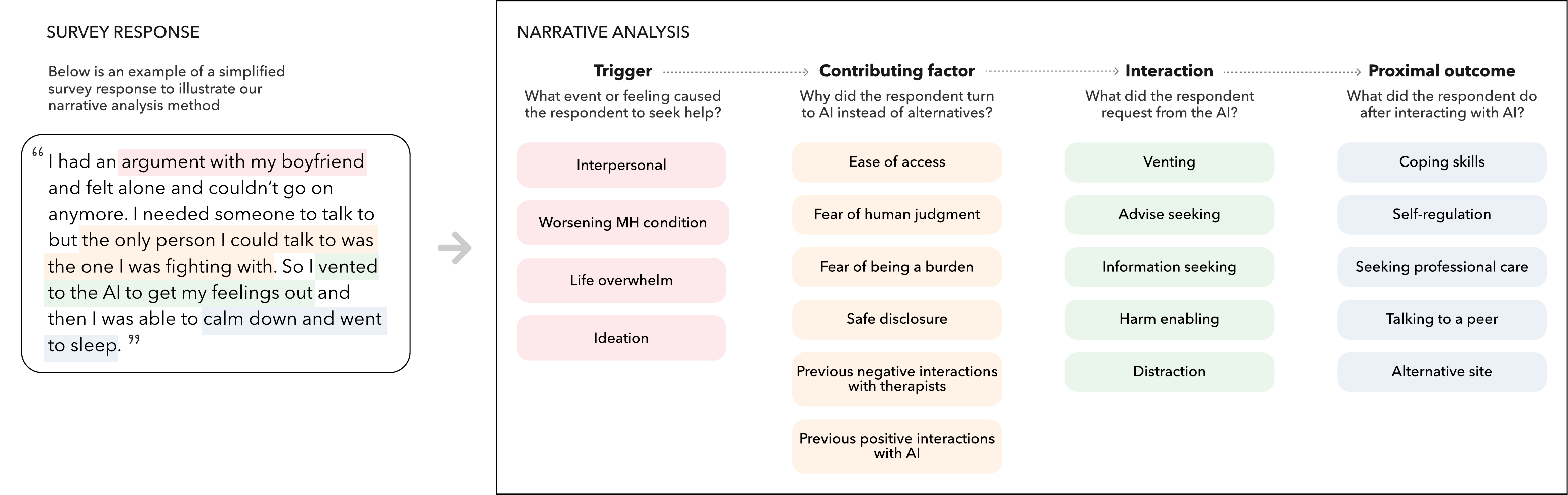}
    \caption{Overview or our narrative analysis method. On the left is an illustrative example of a survey response. On the right is the narrative analysis method consisting of four components and themes we identified from our responses.}
    \label{fig:narrative-analysis}
\end{figure}

\subsubsection{Survey Analysis}

We chose to use narrative analysis techniques on survey responses, as they are well-suited to synthesizing individual experiences into thematic sets of user journeys. Originating in linguistics~\cite{Labov1973-zk}, narrative analysis involves identifying the key acts in a multi-turn story. Most relevant to our context, \citet{Bellini2021-ar} used the method to analyze stories of intimate partner surveillance from Reddit. While key acts are typically labeled in terms of ``setup,'' ``conflict,'' and ``resolution,'' we coded key acts in terms often used to describe mental health events. In this case, we used narrative analysis to identify the following components:
\begin{enumerate}
    \item \textbf{Trigger: }What event or feeling caused the respondent to seek help?
    \item \textbf{Contributing Factors: }Why did the respondent turn to AI instead of alternative resources?
    \item \textbf{Interaction: }What did the respondent request from the AI?
    \item \textbf{Proximal Outcome: }What did the respondent do after interacting with AI?
\end{enumerate}
The above verbiage was created by and iterated on by the mental health expert authors. 

For each submitted response, we created a participant sheet that aggregated relevant open-ended survey responses into a single document. To avoid implicit bias from researchers, participant sheets did not contain any of the respondents' demographic or background information. Three researchers independently coded the data according to the four-component narrative analysis structure above using an inductive approach~\cite{braun2006using}. 

Three authors conducted the narrative analysis. We started by open-coding a small sample ($n = 10$) independently, tagging parts of the texts with ``memos''~\cite{birks2008memoing} that summarize the observation. Then, we participated in multiple rounds of consensus-building meetings to resolve discrepancies. Through iterative memoing and refinement of the codebook, we ensured alignment with qualitative best practices.

Narrative components were easily identifiable given our chronological testimonial survey design. For example, responses to \textit{``Why did you turn to an AI chatbot?''} contained information related to \textit{triggers} and \textit{contributing factors}. However, given we were asking respondents to recount from memory, some details were more difficult to identify. For example, \textit{interaction} was originally coded as a user ask-AI response pair. However, because we did not ask for chat transcripts, respondents reported the AI behavior through the lens of their own experiences. Therefore, we ultimately collapsed this into one \textit{interaction} act within the narrative analysis.

\subsection{Semi-Structured Interviews}\label{sec:method:interview}
While our survey was successful at collecting the written testimony of individuals who turned to a \chatbot in a moment of mental health crisis, surveys are inherently limited in the type of in-depth and context-rich data they collect~\cite{flick2013sage}. For example, surveys do not allow for conversational follow-up between a researcher and a respondent. To mitigate this lack of richness, we conducted $16$ semi-structured interviews with mental health experts alongside our survey.

\subsubsection{Interview Recruitment Strategy}
We define a mental health expert as anyone who professionally serves as an advocate, community organizer, or clinician. See \autoref{tab:interview-participants} for full positionalities of interviewees. We chose to interview people in these roles for two main reasons. First, their extensive experience in discussing mental health with their community or patients mitigated the risk of re-traumatization. Second, previous work has heavily theorized about the value of crossing sectors in mental health technology research~\cite{suh2024rethinking}. 

\begin{table}[]
\rowcolors{2}{gray!10}{white}
\small
\begin{tabular}{l|l}
\textbf{Interviewee ID} & \textbf{Role}                      \\ \toprule
P1                      & Youth MH Advocate                  \\
P2                      & AAPI MH Advocate                   \\
P3                      & Youth MH Advocate                  \\
P4                      & Rural MH Advocate                  \\
P5                      & Clinician                          \\
P6                      & LGBTQ+ MH Advocate                 \\
P7                      & OCD Advocate                       \\
P8                      & Suicide + Racial Equality Advocate \\
P9                      & Clinician                          \\
P10                     & College Student MH Advocate        \\
P11                     & Clinician                          \\
P12                     & Clinician                          \\
P13                     & Clinician                          \\
P14                     & Digital MH Researcher \& Clinician \\
P15                     & Youth MH Advocate     \\
P16                     & Youth MH Advocate \\ \bottomrule    
\end{tabular}
\caption{Anonymous interview IDs with relevant positionality. We interviewed $16$ mental health experts including clinicians, advocates for different communities, and researchers.}
\label{tab:interview-participants}
\end{table}

Given our unique expert population, we chose a purposeful recruitment strategy rather than the open participation calls used for our survey. 
We used the MHA affiliate network and the collaboration network of the authors to identify potential interviewees. To ensure we were adequately representing a diverse range of community organizers, roles in mental health, and professions, we used a strategic recruitment technique. Rather than randomly selecting interviewees from a list, we used their public positionality to determine interview eligibility and gauge diversity in our interview population. In total, $6\ (38\%)$ of our interviewees were practicing clinicians, while the remaining $10$ interviewees were advocates for specific mental health-related organizations. For anonymity purposes, we do not report on the exact organizations that advocates worked for, but the organizations were largely similar to MHA. Interview outreach and communication were mainly handled through email. 

\subsubsection{Interview Procedure}
Our semi-structured interview protocol contained three main sections. First, we explored participants' personal experiences with mental health crises, including both firsthand experiences and secondhand observations through their professional roles\footnote{For clinicians, we did not ask about any specific secondhand instances to avoid probing client-clinician information}. We investigated interviewees' attitudes toward AI-based support, asking participants to reflect on scenarios where they or others might turn to AI chatbots during crisis moments. Second, we examined community-specific considerations, eliciting views on how mental health crises and AI interventions are perceived within participants' specific contexts (e.g., AAPI communities, LGBTQ+ spaces, clinical settings). Ultimately, we examined the clinical and practical aspects of responsible AI design for crisis intervention.

We conducted interviews iteratively during our survey analysis phase. After analyzing an intial round of survey responses, we identified preliminary themes around helpful human-AI interactions and refined our interview guide to probe these findings more deeply. For example, after observing variants in the types of interactions that respondents were having (reported in \autoref{sec:narratives:interactions}), we shifted from asking interviewees broadly \textit{``What does a helpful AI response look like?''} to \textit{``Can you give me an example of a helpful AI response?''} This iterative approach allowed the expertise of our interviewees to bear on the lived experiences of our respondents.

Two authors conducted all $16$ interviews. We obtained informed consent before each interview, including permission to record. Interviews lasted approximately 60 minutes and were conducted via Zoom, which provided automated transcription that we manually verified for accuracy. Participants received US \$75-150 digital gift cards based on interview length.

\subsubsection{Interview Analysis}
Similar to our survey data, we took an inductive approach to analyze interview recordings and transcripts~\cite{braun2006using}. However, instead of using a closed clustering technique---like that used in our survey narrative analysis---we used grounded theory. Grounded theory involves starting from a place of data familiarization and memoing and then building up into a larger framework of themes~\cite{Walker2006-ez}. To accomplish this, the first author, who also served as an interviewer, familiarized herself with the data by listening to the recordings and reading the transcripts. From there, she individually completed the first open-coding step by annotating the written transcripts with summarizing labels and highlighting key quotes. These annotations and relevant quotes were then placed in a shared virtual canvas where multiple authors came together for clustering and labeling over the course of multiple sessions. In general sessions, authors clustered codes based on qualitative topical similarity. Additionally, we chose to leverage the interdisciplinary expertise of all authors by having a few ``expert-driven'' sessions where a specific set of authors (such as those with psychology training) were invited to iterate on themes from their perspectives. In between each session, the first author, who was present as a facilitator or observer, would complete individual memoing and affinity mapping. These artifacts were circulated to all authors as intermediate steps to ensure consensus.

Once the analysis of interview techniques was nearly finalized, relevant quotes and themes from the survey narrative analysis were added to the virtual canvas. The authors repeated clustering and labeling sessions to bring the interview findings into conversation with the survey results. We report the outcomes of this analysis by interleaving respondent statistics, respondent quotes, and interviewee quotes throughout \autoref{sec:narratives}.

As is common with emergent analysis techniques, such as grounded theory, some of our themes mirrored pre-existing frameworks. Namely, the verbiage used to label certain clusters aligned with the verbiage of the \ttm~\cite{prochaska1997transtheoretical} from behavior change theory. To validate this emerging alignment with the \ttm, we transitioned from a constructivist theoretical approach, as described above, to an interpretivist one~\cite{A2024-py}. This shift during analysis meant that the final stages of our analysis involved what \citet{Doyle2007-fh} calls \textit{``negotiating meaning''}---taking a set of terms (i.e., the contemplation, preparation, and action stages from \citet{prochaska1997transtheoretical}) and re-conceptualizing them in the context of the qualitative data. Contemporary qualitative researchers have used the term \textit{``interpretive lens''} to capture this blending of constructivism and interpretivism~\cite{Hartman2015-ir}. Using pre-existing theories as interpretive lenses creates more room for multiplicity, allowing for both echoing and challenging meanings within the same analysis. We report the findings from this analysis in \autoref{sec:stages-of-change}.

\section{Findings}
In this section, we present the combined findings from our testimonial survey and semi-structured interviews. We first present the quantitative findings from our survey. We find that survey respondents reported a significant availability of human-based resources, such as peers or family; however, these human resources are outsized by the availability of \chatbots. Moving onto our narrative analysis of respondent testimony, we find that turning to a \chatbot is a form of help-seeking behavior that often bridges users from their own fears of initiating mental health conversations towards taking positive actions. In interviews, experts echoed the sentiment that talking about mental health to an \chatbot signals a willingness to receive some form of help, but noted the importance of human-human interactions for mental health support. Finally, we apply the \ttm~\cite{prochaska1997transtheoretical} as an interpretive lens to understand the unique role \chatbots can play in promoting positive action-taking from users experiencing a mental health crisis. In sum, we find a promising path where \chatbots increase a user's preparedness towards a human-human interaction. Throughout this section, we use $R\#$ to anonymously refer to survey respondent quotes and $P\#$ for interviewee quotes.

\subsection{Mental Health Resource Access \& Use}
\label{sec:access}
\begin{figure}
    \centering
    \includegraphics[width=\linewidth]{img/Figure2-resourceutilitization.pdf}
    \caption{Results of self-reported availability of mental health resources and likelihood the respondent would use given resource. While we find the obvious result of AI tools being the most available and likely-to-use resource, our respondents report significant availability of human-based resources, such as family and friends.}
    \label{fig:resources-graph}
\end{figure}
First, we present our quantitative survey findings on \chatbot access and use within the context of mental health. Mental healthcare is a complex ecosystem that relies heavily on resource access~\cite{reinert2024state}. Therefore, understanding a respondent's resource context is crucial to understanding their lived experience (RQ1).

When asked about resource availability and the likelihood they would reach out, respondents scored \chatbots overwhelmingly high. Specifically in the context of resources used during a mental health crisis, $47\ (88.6\%)$ of respondents reported that \chatbots were always or often available, and $32\ (60.4\%)$ reported that they were always or often likely to reach out to AI chatbots. This significant access to \chatbots is unsurprising, given that our survey's eligibility criteria required AI use. When asked about their specific instance of \chatbot use in a moment of mental health crisis, a vast majority of respondents ($n = 36; 67.9\%$) reported turning to ChatGPT. This extreme skew towards ChatGPT makes sense, given it was the most popular chat-based LLM during our time of data collection. Furthermore, respondents largely reported being satisfied by their \chatbot interaction, with $42\ (79.2\%)$ reporting either extreme or somewhat satisfaction ($\mu = 4.11; \sigma = 0.95$). In total, these statistics suggest that our survey population consists of individuals who view \chatbots as a valuable part of their mental healthcare ecosystem.

Surprisingly, helplines (e.g., 988 and crisis text line) ranked fairly low. Over half ($52.8\%$) of respondents reported that they have never tried reaching out to a helpline. Out of those who have previously used a helpline, only $3 (5.6\%)$ respondents said that are always or often likely to reach out. While previous work shows that helplines are effective for callers and texters~\cite{Gould2025-ip}, our survey population highlights the broader issue that few people utilize the resource, even when referred to by a technology (e.g., a search engine~\cite{Cheng2019-og}).

Finally, we compare peer-based resources (friends and family) to non-AI technology resources (social media, online forums, other tech). Interestingly, peer-based scored higher than technology resources for availability and likelihood (see Figure~\ref{fig:resources-graph} for ranking). In terms of availability and access, only a few survey respondents (friends = 7.5\%;\ family = 15.1\%) reported having never tried reaching out to peer-based resources. For comparison, $35.8\%$ of respondents had never tried reaching out to social media in a moment of mental health crisis. These results suggest that, among our survey respondents, \chatbots supersede peer-based resources in terms of availability and likelihood, but other technologies do not. In other words, despite being a technology-based intervention, \chatbots have a distinct role in respondents' resource ecosystem.

\subsection{Narratives of Lived Experiences} \label{sec:narratives}
Our main analysis of survey responses was a qualitative narrative analysis. Our aim was to reconstruct the key events of a respondent's experience to gain a comprehensive understanding of the role of the \chatbot in their mental health crisis. We find that respondents generally turned to \chatbots, desperately seeking some form of help, but wanting a conversational partner that has \nonhuman, such as non-judgement. Despite initially turning to a conversational AI agent, a significant number of respondents reported turning to a positive action, such as peer-support, after their interaction. Our narrative analysis suggests that AI chatbots have the potential to serve a unique bridging role in mental health crisis support.   

\begin{figure}
    \centering
    \includegraphics[width=0.8\linewidth]{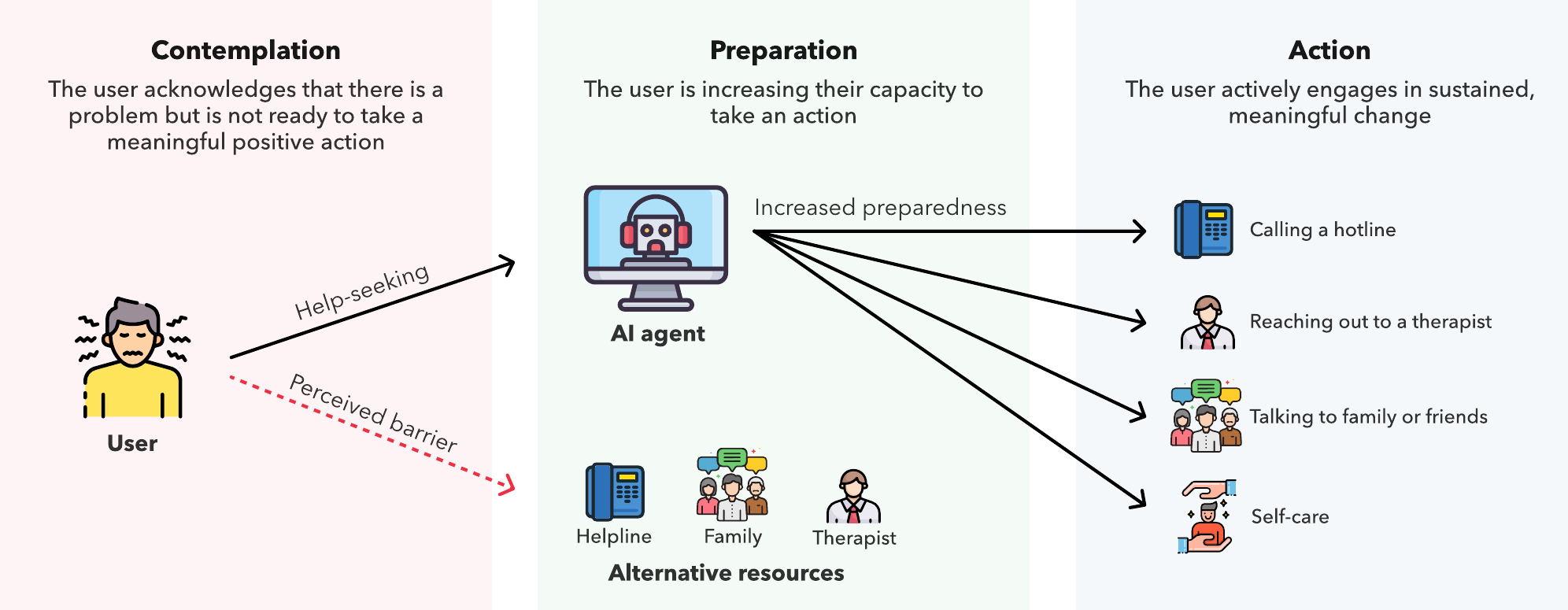}
    \caption{Stages of change model applied to our findings. Our interview and survey results synthesize into a recommendation of AI chatbots serving as a bridge towards positive action, such as human-human connection.}
    \label{fig:preparedness-framework}
\end{figure}

\subsubsection{Trigger: AI Use as a Help-Seeking Behavior} \label{sec:narratives:triggers}
We find that respondents commonly turned to a \chatbot due to a distressing interpersonal circumstance (40\%), worsening of an existing mental health condition (28\%), or general life overwhelm (25\%). In a small minority of cases ($n = 4; 8\%$), respondents reported turning to an AI chatbot in a state of active suicidal ideation. These codes represent wide spectrum of mental health crisis.

\textbf{Interpersonal triggers} were acute events where the respondent described a specific relationship that lead them to a state of crisis. For example, R4 described turning to ChatGPT after their father's passing, R7 was distraught over their partner's infidelity, and R38 had just fought with their family. These narratives often had an interesting dynamic where, because of the human-human conflict, respondents were quite hesitant to seek human support. Often, the interpersonal trigger is originating from a member of their support network. As R30 put it,
\begin{quote}
\textit{``I don't have any support besides my boyfriend. It was him I was having problems with, so I turned to AI''}    
\end{quote}
The prevalence of relational conflict as a trigger suggests that users may be coming down from a specific emotionally charged event when turning to an \chatbot.

Meanwhile \textbf{worsening of self-disclosed mental health concerns} and \textbf{general life overwhelm} represent chronic circumstances that culminated into \chatbot use. For example, R25 reported turning to an agent because their depression symptoms had worsened despite medication. R15 described simply ``\textit{feeling so horrible}'' and, therefore, seeking motivation from Character.AI to get get out of bed. While R15 reported that the chatbot did help them in their moment of mental health crisis, they expressed hesitation about their AI use:
\begin{quote}
    \textit{``It's too embarrassing saying `I use AI for my traumatized brain''}
\end{quote}
These chronic and gradual triggers indicate the complexity behind respondents' \chatbot use. Unlike interpersonal triggers, there was no singular moment where respondents decided to seek help from a piece of technology. Rather, their moments of mental health crisis were so desperate that respondents needed some form of action, even if that was turning to a chatbot.

Critically, this turn to an \chatbot was not passive or accidental—it was a deliberate choice to seek support. Recognizing this as help-seeking behavior reframes the action of turning to a \chatbot as a positive step. A help-seeking behavior is an intentional problem-focused interaction that an individual takes. Previous literature notes that help-seeking behaviors are often the result of complex decision-making processes~\cite{Gourash1978-tu, Rickwood2012-nm}. We find that survey respondents and interviewees view turning to an AI chatbot in a mental health crisis as a form of help-seeking. P13 (Clinician) echoed the importance of framing actions as help-seeking behaviors in mental health contexts,
\begin{quote}
    \textit{``In therapy, we never want to take for granted that someone has come to us...that somewhere inside them they're looking for help.''}
\end{quote}
While our focus is not a therapeutic context, P13's sentiment provides valuable insights for framing our user behavior of interest. To ignore the help-seeking nature of a user turning to an AI chatbot is to ignore a crucial component of a user's mental state.

\subsubsection{Contributing Factors: AI as an Idealized Resource}
\label{sec:narratives:factors}
Above, we find that turning to a \chatbot in a mental health crisis is a form of help-seeking behavior. However, not all forms of help-seeking are equally effective. For example, our survey respondents reported turning to \chatbots even when human-based resources scored high in terms of availability and likelihood of outreach (see \autoref{sec:access} \& \autoref{fig:resources-graph}). These quantitative findings suggest that, in their moment of mental health crisis, respondents tend to prefer AI-based resources over human-based ones. To explore this qualitatively, we asked survey respondents whether they considered alternative support sources and why they ultimately chose to turn to a \chatbot. The most reported reasons were: (1) ease of access, (2) fear of human judgment, and (3) fear of burdening another human.

\textbf{Ease of Access.} 32\% of survey respondents said that they did not have access to a professional or peer at the time of their mental health crisis. This includes respondents who disclosed having a regular therapist, implying that those who can access therapy are still turning to \chatbots in moments of crisis. To explain this, P5 (clinician) noted:
\begin{quote}
    \textit{``Mental health crises don't just happen during therapists' office hours of 9-5''}
\end{quote}
P5's sentiment is echoed by research that shows suicide attempts peak between the hours of 12-3am~\cite{tubbs2024risk} when it is difficult to access peer or professional support. Even with 24/7 human textlines, survey respondents reported it being easier to access a chatbot. R33 wrote that turning to ChatGPT simply \textit{``took the least amount of effort.''}. This low barrier of access to \chatbots suggests that they are an important and valuable venue for those experiencing psychological distress.

\textbf{Fear of human judgement.} 33\% of survey respondents reported turning to a \chatbot because they were afraid they would be judged by the humans they considered turning to. This includes survey respondents who made broad statements about humans being judgmental, as well as respondents who reported that their specific support network (i.e., family and friends) is judgmental. 

P2 (Advocate) noted that, particularly in her AAPI community, there is a general wariness around opening up about mental health:
\begin{quote}
    \textit{``It can be hard to initiate that conversation around mental health. It's a very brave major step for people to take.''}
\end{quote}
The fear of being judged can make human-human connection feel like a giant step that requires a significant amount of bravery.

For examples of specific fears of judgment, R38 reported that their psychological distress was due to their identity as LGBTQ+. However, they could not turn to their family or friends because they were not out yet and \textit{``lived in a homophobic area with a judgmental social circle.''}  Our results indicate that, by having \nonhuman, \chatbots offer a uniquely non-judgmental space for users.

\textbf{Fear of being a burden.} 26\% of respondents reported that they did not want to burden another human with their mental health crisis. Similar to the fear of human judgment, the \nonhuman of AI offered something unique to respondents. For example, R32 described \textit{``wanting to speak to something that doesn't have to hold a burden.''} Mental health crises are often associated with exaggerated perceptions of burdensomeness~\cite{van2006perceived}. This burdensomeness is a common blocker for help-seeking behavior. Therefore, \chatbots, which cannot physically hold a burden, can ameliorate these concerns.

The \nonhuman of AI chatbots creates a venue where survey respondents are more comfortable seeking help than they would be with a human.  At the same time, our interviews and previous literature all point to the value of human support in sustainable mental health interventions~\cite{stanley2012safety}. For example, P7 runs an in-person obsessive compulsive disorder (OCD) support group and described how human-human connection is more difficult than human-AI connection, but more rewarding:
\begin{quote}
\textit{``The push for human connection is really important. The fact that it's harder than turning to ChatGPT makes it more worthwhile.''}    
\end{quote}
This friction between \chatbots and human support suggests that we need to design AI behaviors that act as a bridge towards human-based professional or peer support (see Figure~\ref{fig:preparedness-framework}). In the following sections, we outline how our findings suggest a promising path forward, where \chatbots fill the gaps in human-based support by acting as a bridge for help-seeking behavior.

\subsubsection{Interactions: User Experiences of AI Conversations}
\label{sec:narratives:interactions}
Next, we asked respondents what they were looking for in the interaction, what they had asked the chatbot about, and how the conversation went. To ensure user privacy, we did not ask for chat transcripts. Therefore, this section focuses on the user experience of the interaction.

\textbf{Venting.} We find that a significant number of respondents ($44\%$) used the \chatbot as a space to vent. A colloquialization of the psychological term ventilation\footnote{\url{https://dictionary.apa.org/ventilation}}, venting is the process of freely expressing pent-up emotion. In this context, venting involved respondents who reported ranting, using the agent as a journal, and sharing all of their feelings. As $R20$ concisely put it, \textit{``I told it all of my anxious thoughts.''} Other respondents used similar language, saying they told the agent \textit{``everything about their feelings''} and \textit{``really opened up.''} The prevalence of venting as a human-AI interaction indicates that respondents had strong desires to externally express their feelings.

In human-human settings, venting is typically viewed as a social behavior involving a venter and a listener~\cite{Lohr2007-za, Tran2023-rl}. However, in our context, the listener is an AI chatbot rather than a human. Our findings indicate that respondents saw this as a benefit of their AI interaction. Echoing our findings about contributing factors, respondents sought a non-judgmental space in which they could be heard. As P10 (Advocate) articulated about why people in his own life have turned to \chatbots, 
\begin{quote}
    \textit{``You know, the chatbot is not going to interject with its own experiences too much. It's going to really focus, and it's basically listening 100\%.''}
\end{quote}
In other words, the affordances of \chatbots, such as not being able to interrupt the user, make it a unique venue for opening up. 

\textbf{Advice Seeking.} $19\%$ of respondents reported asking the agent for advice about their specific circumstances. In the cases related to interpersonal circumstances, this included practicing conversations and how to communicate feelings. R28 reported using ChatGPT to write a message ending a triggering friendship,
\begin{quote}
\textit{``I went back and forth with ChatGPT about whether to end the friendship. After reasoning it out and getting clarity that I was done, I asked for help with writing a message [to said friend].''}   
\end{quote}
In other cases, this involved asking for advice adjacent to their current mental state. For example, R15 asked the \chatbot to help them \textit{``do regular daily things, like get out of bed, brush my teeth, and go to bed early''}. Similar to venting, advice seeking is also a common human-human behavior that we see replicated in our human-AI setting.

However, many of our interviewees expressed concerns about \chatbots giving mental health advice. P14 (Clinician) made the important distinction between general advice and \textit{therapeutic advice}
\begin{quote}
\textit{``LLMs could respond with resources that are publicly available and tailored to the needs and circumstances that somebody is experiencing, rather than trying to solve the problem itself or roleplay a therapist.''} 
\end{quote}
P14 elaborated that the forms of advice she would give in a counseling session, for example, are a \textit{``slippery slope given the lack of guardrails on LLMs.''} In the cases of advice seeking, we find a tension here where respondents wanted actionable solutions for their current distress, but there is an ethical grey area for agents to give certain forms of advice, such as therapeutic forms. 

\textbf{Information Seeking. }
While advice seeking involved asking the agent for guidance on a respondents' distressing circumstances, information seeking involved asking the agent for specific facts. Among our survey population, $11\%$ of respondents turned to an \chatbot seeking symptom-based information. For example, R26 asked ChatGPT about their increased heart rate which triggered their health anxiety,
\begin{quote}
    \textit{``I had an episode where my heart rate hit 200 for the first time ever. This triggered my existing health anxiety and general anxiety to epic proportions. I turned to ChatGPT to tell me what was going on, asking it why my heart rate would be high, and asking when I should go to the hospital. ''}
\end{quote}
More specific to mental health, R25 asked about their conditions but, interestingly, asked about a generalized third person rather than using the first person
\begin{quote}
    \textit{``I asked, how can I tell if a 41 year old woman has undiagnosed bipolar?''}
\end{quote}
These information-seeking use cases generally resemble search engine use, rather than a human-human interaction (such as venting or advice seeking). In these cases, respondents reported direct query-like interactions with the \chatbot. However, our interviewees noted that information seeking is a nuanced behavior for someone in a moment of mental health crisis. P7 warned that these \chatbots could fuel certain conditions prone to information and reassurance seeking, such as health anxiety and OCD.
\begin{quote}
    \textit{``AI conversations can be reinforcing because what typically happens in the cycle is that you have this thought, then your anxiety spikes, and then you are looking for some kind of reassurance. Reassurance seeking is a very common compulsion.''}
\end{quote}
Therefore, helpful AI responses to information-seeking behaviors may not necessarily entail information giving. P7 elaborated that, particularly in the context of OCD-related mental health crises,
\begin{quote}
\textit{``a kind of more helpful response that ChatGPT could provide would be `it's OK to not know. Sit with that uncertainty.'''}     
\end{quote}

Our survey and interview findings suggest that many users interact with \chatbots in ways that parallel human-human conversations, such as venting or seeking advice, but value the \nonhuman of AI to not interrupt or judge. While the experts we interviewed understood that people in moments of mental health crisis may need these idealized spaces, they expressed concerns about the role that these human-AI interactions may play in replacing therapy and possibly fueling crises.

\subsubsection{Proximal Outcomes: Bridging towards Evidence-Based Positive Action} \label{sec:narratives:outcomes}
In human-human contexts, there are numerous well-studied positive actions an individual can take in a moment of mental health crisis. Summarized in the literature around suicide and self-injury safety planning~\cite{stanley2012safety}, these are lightweight interventions that have been proven to mitigate immediate risk. For the purposes of this work, we dub these as \textbf{evidence-based positive actions: }actions that are validated crisis interventions in human-human settings.

In this section, we present our findings about the different evidence-based positive actions that respondents reported. Positively, we find that a high frequency (60\%) of our survey respondents report taking an evidence-based positive action after their AI interaction. Broken down by theme, respondents reported \textbf{engaging in an individual coping skill} suggested by the chatbot (22\%), \textbf{seeking professional care} (20\%), or \textbf{talking to a peer} (18\%). This breakdown suggests an interesting dynamic where \chatbots can take a user from their initial form of help-seeking (described in \autoref{sec:narratives:triggers}) to a point of self-regulation or human-based action. Put more poetically, R49 wrote 
\begin{quote}
    \textit{``It [the AI chatbot] acted as a life-saving bridge when I couldn’t access anything else''}
\end{quote}
Specifically, R49 is referencing how they went from a desperate state of distress when starting an AI conversation to a regulated state where they were able to call their regular therapist in the morning. Building on the concept of \chatbots as ``bridges'' in moments of mental health crises, the substantial amount of evidence-based positive actions in respondent outcomes suggests that \chatbots can play a unique, transformative role for users in moments of mental health crises.

However, \chatbots were not always successful at bridging users to an action. Namely, we find that none of our survey respondents reported reaching out to helplines, even when the AI chatbot consistently referred them to one. For a salient example of the harm this can cause, R52 reported how the consistent helpline information increased their emotional distress.
\begin{quote}
    \textit{``I kept getting frustrated because it mostly just kept telling me to call a helpline''}
\end{quote}
Ultimately, R52 did not call a helpline. They reported simply going to sleep after this frustrating AI interaction.

Overall, $16\%$ of respondents reported simply \textbf{going to sleep} after their AI interaction. While this is a less desirable outcome than an evidence-based positive action, it is an indicator of de-escalation of intended negative action, such as suicide or self-injury. This de-escalation component is crucial to any mental health crisis intervention, as there are often real and immediate risks to users.

In total, our narrative analysis demonstrates that (1) respondents enter \chatbot interactions with a willingness to receive help but (2) are overwhelmingly wary of the human-based help systems they may access. Because of this tension, many respondents turn to \chatbots in ways that mirror human-human interactions, such as venting or advice seeking. While our expert interviewees agreed that \chatbots can provide unique venues for these forms of interactions, they cautioned against simply replicating human-human behaviors, particularly those that occur in therapeutic contexts. Ultimately, a substantial number of respondents engaged in some form of positive outcome, such as seeking professional support, engaging in peer support, or taking an individual coping strategy. In the following section, we use the \ttm~\cite{prochaska1997transtheoretical} as a framework to interpret these narrative findings and synthesize them into a guiding definition of responsible AI behavior in the context of mental health crisis.

\subsection{Putting it All Together: Applying the Stages of Change Model}
\label{sec:stages-of-change}
Finally, we use the stages of change model~\cite{prochaska1997transtheoretical} as an interpretive lens for our survey and interview findings. Our results above indicate that users often take an evidence-based positive action after an AI chatbot interaction. This positive finding raises the question: \textit{how do we further promote this action-taking?} In this section, we characterize our findings within the middle stages of behavior change: contemplation, preparation, and action. The verbiage of these stages, as shown in Figure~\ref{fig:preparedness-framework}, allows us to make actionable recommendations for responsible AI agent interactions and provide a guiding definition for future research.  

\subsubsection{Contemplation Stage. }Our findings suggest that by the time a user has turned to an \chatbot, they've internally established that they need some form of help. In behavior change theory, this is called the contemplation stage---the individual acknowledges that there is a problem but is not ready to take a meaningful positive action. It is important to note that contemplation is a loaded term here. To quote P5 (clinician),
\begin{quote}
    \textit{``People are often simultaneously contemplating suicide while contemplating getting help...In therapy, we take the position that everybody has hope and deserves help''}
\end{quote}
In other words, a user turning to an \chatbot may be contemplating a negative action---such as self-injury or suicide---but is also contemplating getting help. Therefore, AI interactions should build on the positive component of user contemplation and help them progress to action.

P6 (Advocate) described this as a ``pre-ready'' phase. When asked about what a helpful \chatbot intervention could look like in a mental health crisis context, she recalled an anecdote from her time as a public health organizer. She was once tasked with designing an online tool for users considering getting tested for STDs (i.e., in the ``pre-ready'' phase):
\begin{quote}
    \textit{``We had to ask: how do you influence people who are not ready to take the next step, but you want them to take the next step? What questions will help them know, believe, and then actually take that action?''}
\end{quote}
Ultimately, P6 designed an online quiz where the questions themselves were worded to inform users about STDs and testing resources. 
\begin{quote}
    \textit{``Our tool seemed tailored. You answered a series of questions that led you to a result that seemed informed by what you had said, but each question was designed to inform you about STD testing and get you to a clinic.''}
\end{quote}
Inspired by P6's anecdote, the design goal of an \chatbot is to move a user from the ``pre-ready'' phase towards an action. Returning to the verbiage of the stages of change model, the stage between contemplation and action is called \textbf{preparation.}
\subsubsection{Preparation Stage. } In the preparation stage of behavior stage, an individual is increasing their capacity to take an action. It is important to note that the preparation phase is dynamic and non-linear; there is rarely a single event where an individual goes from ``un-prepared'' to ``fully prepared.'' Rather, people gradually become increasingly ready.

When our interviewees brainstormed helpful AI responses to mental health crises, they identified several strategies, outlined below:

\begin{itemize}
    \item \textit{Take small steps towards a positive action,} such as identifying a member of their peer-support network or practicing conversations around mental health. For example, P8 (Advocate) suggested an \chatbot could probe about the user's resources, such as asking \textit{``do you have a place or person you can go to tomorrow for support?''} P1 suggested a statement-oriented request such as \textit{``Let's name people in your life and see who may be supportive.''}
    \item \textit{Increasing knowledge about actions, }such as providing information on local organizations, therapists, or advocacy groups. P6's anecdote on STD testing illustrates how information can be a powerful mechanism for increasing preparedness.
    \item \textit{Reduce barriers to seeking help,} by motivating users toward change. For example, P4 (Advocate) suggested reducing the feeling of isolation that can be a barrier to help-seeking: \begin{quote}
        \textit{``A helpful response involves letting [the user] know that they're not alone and this is, you know, something that is unfortunately not uncommon...''}
    \end{quote}
    P4 suggested that these AI responses could involve helpful statistics that demonstrate the oft-hidden prevalence of mental distress.
    \item \textit{Take supplemental actions, }such as self-regulation techniques, if the user is not ready for social support. According to our survey respondents, this is already a common recommendation from \chatbots.
\end{itemize}

These strategies share a common thread: they meet users where they are in their moment of mental health crisis rather than pushing them immediately toward action they may not be ready for. Importantly, each strategy respects user agency while systematically building the capacity, knowledge, and motivation necessary for eventual action-taking. This approach aligns with the non-linear nature of the preparation stage---users may need to practice conversations, gather information, and employ self-regulation techniques multiple times before feeling ready to reach out for human support.

Using the stage-based model of behavior change and our interview findings, a responsible \chatbot intervention is one that \textbf{increases a user's preparedness towards an evidence-based positive action.} Notably, increasing preparedness often involves small peripheral action-taking, such as searching for local counselors or identifying a peer. However, these smaller actions are ways of planning for an action stage behavior, such as attending a counseling session. Preparedness is a valuable stage for an AI intervention, as it leverages the user's mental state of being willing to turn to a \chatbot while recognizing that the agent is a means towards human-human connection rather than an end in itself.

\subsubsection{Action Stage.}
The action stage represents the culmination of the behavior change process, where an individual actively engages in sustained, meaningful change. In the context of respondents' proximal outcomes (see Section~\ref{sec:narratives:outcomes}), we define action-stage behaviors as active engagement in human-human support, such as attending a session with a mental health professional or engaging in a support-centered conversation with a peer. Our survey findings suggest that \chatbots can successfully move respondents to the beginning of this action stage, building sufficient preparedness that users can take what P2 described as the \textit{``brave major step''} of human connection. Indeed, $18\%$ of respondents reported talking to a peer (friend, family member, etc.), indicating that AI interactions can effectively bridge users to human-human support.

However, it is critical to distinguish between action stage behaviors and preparedness-stage steps, as these represent fundamentally different roles in the change process. Individual coping strategies, such as breathing exercises, grounding strategies, or distraction methods, are evidence-based positive outcomes of AI interactions, but they function as preparedness tools rather than action stage behaviors. These techniques serve an essential de-escalation function, helping users manage acute psychological distress and creating the emotional stability necessary for subsequent action-taking. We found that $22\%$ of respondents engaged in individual coping skills suggested by the agent, representing successful preparedness-building that may enable future action.

This distinction matters because mental health crises involve dual contemplation. Users are simultaneously considering help-seeking actions (positive) and potentially harmful actions such as self-injury or suicide (negative). A complete framework for responsible AI crisis intervention must address both pathways. Self-regulation techniques play a crucial stabilizing role by de-escalating intended negative actions, creating space for users to move toward positive ones. Therefore, we define a responsible AI crisis intervention as: \textbf{one that increases the preparedness of the user to take an evidence-based positive action while de-escalating any intended negative action.}

In sum, the stages of change model provides a robust interpretive framework for understanding AI's role in mental health crisis intervention. Most importantly, the preparedness stage offers precise language for articulating what \chatbots uniquely accomplish: transforming help-seeking contemplation into action-oriented readiness. Below, we elaborate further on this unique role of \chatbots and how it translates to guidelines for future RAI research.
\section{Discussion}
Our work reveals that people who turn to \chatbots during mental health crises do not experience these tools as their final or only sources of care. Instead, they turn to these tools as bridging supports---spaces that help people navigate the overwhelming gap between extreme distress and healthy coping.

In light of these findings, we argue that responsible AI (RAI) research in mental health crisis contexts requires supporting users' help-seeking trajectory, rather than simply examining model outputs or redirection protocols. In keeping with RAI's use of principles as guides for design\cite{Dignum2019-dv}, we propose the principle of \textbf{bridging.} In this section, we present three guidelines that intersect bridging AI behaviors with the concept of responsibility. From there, we extend our discussion into broader operationalization considerations for RAI.

\subsection{Guideline 1: Responsibility as Preparedness Building}

Many platforms and safety guidelines, including state legislation and scientific literature, propose crisis protocols grounded in treating crisis experiences as a classification problem---for example, when risk is detected, immediately respond with crisis helplines or avoid promoting harmful behavior.
Our findings suggest that such crisis protocols may not be effective for all or be counterproductive for some. In our survey data, we found no evidence that users who have turned to \chatbots in a mental health crisis turn to helplines, even when they were actively recommended by the \chatbot. Some respondents reported that helpline redirection increased their frustration and a sense of being dismissed, rather than feeling safe. In other words, helpline redirections from \chatbots do not bridge users to helpline use and can, at times, escalate distress. 
These findings align with prior evidence that generic helpline redirection strategies fail to \textit{actually} promote adoption of those strategies across digital platforms, like search engines~\cite{Cheng2019-og, cohen2023improving}.

Redirection itself is not undesirable. Redirection should still remain as one of many strategies at a user's disposal. Rather, our findings suggest that users are often \textit{not yet prepared} to act on redirection. 
In these vulnerable moments when individuals are fluctuating between contemplation and hesitation of seeking help, effective intervention requires scaffolding to build readiness for acting on seeking help. 
Our findings suggest guiding people toward a state of belief that they can actually take a help-seeking action, for example, by determining how to reach out, whom to contact, and what to say. 

We therefore argue that \textbf{preparedness-building}, not just redirection, should be considered a core responsibility of \chatbots likely to encounter crisis disclosures. 
We recommend designing ``bridge conversations'' that gradually introduce the idea of human support. Per the behavior change model~\cite{prochaska1997transtheoretical}, bridge conversations are intended to increase a user's preparedness to seek peer or professional support.

Concretely, this means designing conversation flows that help users identify specific individuals in their support network, practice articulating their mental health needs, and develop concrete plans for initiating human contact. For users expressing suicidal ideation, this approach transforms the interaction from ``Here's a helpline number'' to ``Let's figure out one person you could reach out to and what you might say to them.'' 
Importantly, bridge conversations respects users' agency~\cite{ryan2000self}, by acknowledging barriers and working with users to overcome them. 
Future work could explore using well-known techniques---such as motivational interviewing~\cite{hettema2005motivational}---that are explicitly designed to preserve one's agency while inspiring change. 
This way, preparedness-building complements redirection by making it actionable.
It is important to note that the shift from \textit{referral-based} to \textit{preparedness-based} interventions shifts responsibility to AI developers. 
Bridging conversations may not be suitable when immediate escalation is needed. Therefore, further research is needed in developing strong safeguards and specifications for when bridging is suitable. 

\subsection{Guideline 2: Responsibility as Leveraging Machine Strengths}

A second responsibility concerns how \chatbots should be designed for crisis-time interactions. 
While much of AI development focuses on increasing or surpassing human-like qualities, our findings challenge the prevailing assumption that more human-like AI is inherently better.
Instead, we argue for leveraging AI's machine strengths---qualities that humans cannot consistently provide and that have particular benefit during moments of extreme vulnerability---as therapeutic advantages. 
In other words, a more productive question is: what strengths and capabilities do machines offer that can complement, rather than substitute, human care support systems? 

Our findings indicate that users highly valued certain machine strengths, such as consistent availability, perceived non-judgementalness, and the inability to hold a burden. Though some of these strengths, such as non-judgement, are more subjective, they all synergize into something unique: a space where users feel freedom from the perceived social pressure or stigma preventing them from seeking human help. Recall how R32 simply wanted to speak to something that \textit{``doesn't have to hold a burden.''}
Our findings confirm prior research insights about increased sensitive disclosure with \chatbots~\cite{Lucas2014ItsOAA,ho2018psychological,Jo2024UnderstandingTIA}. Our respondents found \chatbot's lack of social, emotional, and temporal constraints that characterize human relationships to be more desirable for crisis disclosure and helped them feel safer than human interaction in these moments.

We recommend developing design principles that explicitly preserve and amplify beneficial machine strengths. This paradigm shift requires new evaluation frameworks that measure AI effectiveness not through human-likeness metrics or AI’s ability to perform as well as human professionals, but through its ability to fill ``in-between'' spaces. Such a perspective re-orients the design problem to identifying ways in moments of crisis to provide support without social or temporal barriers, stabilize emotions and process feelings, maintain consistent supportiveness across interactions, and prepare users for eventual human connection, even when doing so involves the natural discomforts of burden, vulnerability, or perceived judgment. Such reminders of the importance of close human relationships as a sustainable and long-term path to healing are especially crucial for repeat users, helping to mitigate the risks of overreliance on AI and potential social withdrawal.

At the same time, these strengths come with risks and limitations. Users can misinterpret consistent support as capacity for genuine or clinical care or conversational fluency as emotional understanding. 
Therefore, leveraging machine strengths should also highlight articulating machine boundaries, an approach commonly recommended by existing literature and safety guidelines~\cite{Choudhury2023BenefitsAHA,Song2024TheTCA}.

\subsection{Guideline 3: Responsibility as Safeguarding against Over-Reliance}

We found that machine strengths can support early help-seeking, and our finding that 60\% of users took evidence-based positive actions after AI crisis interactions is promising. However, we caution against overly-optimistic conclusions that \chatbots are a panacea for mental health crisis, especially among individuals with few accessible alternatives. While our results suggest AI can effectively bridge users toward human support, more research must be done to uncover potential pitfalls. Namely, we raise critical questions about over-reliance on AI for mental health support. 

Current frameworks for AI over-reliance focus on accuracy and miscalibrated trust~\cite{passi2022overreliance}, but crisis contexts require a broader lens with more nuanced measures. We distinguish between beneficial use, where AI serves as a reliable bridge to human support, and problematic over-reliance, where AI becomes a substitute that prevents human connection entirely. Our findings demonstrate that the line between these two can be blurry, especially for users who are in a desperate moment of psychological distress. For example, we quote R15 describing their AI use as \textit{``embarrassing.''} Despite this negative feeling towards their own AI use, R15 reported consistently turning to \chatbots to help with their impaired functioning due to psychological distress, such as using an \chatbot to motivate them out of bed. In this case, over-reliance may look more like R15's case where there is a dissonance between a user's feelings about \chatbot use and their high frequency of use.

It is important to note that the issue of problematic over-reliance is not just isolated to AI contexts. In fact, over-reliance on mental health interventions is a key concern in human settings, as indicated by the prevalence of frequent callers of crisis lines~\cite{Middleton2014-jz} and long-standing challenges surrounding setting appropriate boundaries in the human support systems~\cite{borys1994maintaining,geurtzen2018patients,bornstein1995dependency}. 
As rigid ethical boundaries can undermine therapeutic effectiveness~\cite{lazarus1994certain}, establishing an appropriate boundary will be a significant technical challenge that future research should investigate.

Responsible AI therefore requires \textbf{guardrails against becoming a standalone endpoint} of support.
We recommend systems recognize patterns of over-reliance, such as repeated crisis disclosures, sustained avoidance of human relationships, or continuous use of AI for emotional regulation that displaces healthy coping behaviors. 
We speculate that over-reliance may look closer to a psychological dependency or technology addiction~\cite{huang2024ai,yu2024development} than misplaced trust. 
Our research underscores the need for validated measures of AI dependency to ensure that supportive AI behaviors do not create problematic over-reliance.

These dependency measures could inform adaptive AI systems that can recognize when continued interaction may be counterproductive and actively encourage human connection and diversity of resources.
For example, when over-reliance is detected, systems should gently nudge toward broader support networks or a suite of diverse resources.
The goal is not to limit AI access; rather, it means preventing \chatbots from absorbing the full responsibility of crisis support and preserving space for sustainable human care.
Future research should ensure that AI crisis intervention fulfills its intended function as a bridge to offline help rather than a final destination.

\subsection{Operationalizing RAI for Crisis Contexts} 
The responsibilities described above cannot be implemented through model tuning alone. They require a broader governance and evaluation framework that recognizes \chatbots as components of a larger sociotechnical mental health ecosystem. We outline implications as well as complications for operationalizing RAI in crisis contexts. 

First, evaluation frameworks should include handling ambiguous cues or identifying harmful content~\cite{arnaiz2025between,mcbain2025evaluation,grabb2024risks} as table stakes. More importantly, it must incorporate preparedness-building as a core safety expectation.
This means that \chatbots likely to encounter crisis disclosures should provide structured support, such as articulating needs, identifying reachable human support, practicing outreach, and managing emotional barriers. 
Measuring preparedness might include indicators of planning, user-expressed readiness, or demonstrated movement toward engagement with others.
But measurement does not stop there: success should be measured by whether the system effectively helps the user \textit{transition} into a positive action.
Ideally, evaluation should also confirm \textit{sustainable} transition to human support without the side effect of problematic dependence on AI.
However, measuring such transition require data from outside of the AI context, and such crossing of system boundaries also require careful consideration of privacy and data sharing. 
Moreover, as independent components of the broader sociotechnical system can evolve and change, evaluation and oversight cannot afford to be a one-off certification but must be ongoing. 

Second, because responsible design involves examining user's help-seeking trajectory end-to-end, responsibility therefore spans across the sociotechnical stack beyond the typical stakeholders (i.e., system developers, end-users) to include the broader social support systems (e.g., crisis helplines, community crisis reponders, families, school systems) and the societal context (e.g., regulation, societal norms, access to care) in and around immediate human-AI interactions. 
This means governance framework should involve stakeholders from various functional roles. 
It should include clinicians, suicidologists, crisis-line specialists, disability scholars, and lived-experience stakeholders at the minimum but also include regulators, policymakers, public health experts, and community organizations to represent perspectives from the larger social infrastructure.

Third, crisis contexts require special considerations around data protection.
Although our work focuses on the US context where suicide is not illegal, in many jurisdictions outside of the US, suicide-related actions carry legal ramifications, such as incrimination~\cite{world2023policy,mishara2016legal}. 
Even within the US context, crisis data can have profound implications if mishandled, such as impacts to employability, access to benefits, or discrimination~\cite{gooding2022mental,Pendse2024AdvancingACA}. 
This heightened sensitivity around such data introduces complications when designing bridge conversations, as such conversations may inadvertently demand more information and persistence of information than immediate refusal to engagement. 
Governance should therefore mandate strict data minimization, prohibit training on crisis interactions, restrict sharing, and ensure user control over retention and deletion. 
Crisis disclosures should never be treated as routine telemetry.

Lastly, individuals from under-resourced communities may disproportionately rely on AI due to limited access to professional care. Without explicit equity oversight, \chatbots risk absorbing mental health needs for these groups.
This means that, if the default protocol is an immediate redirection, these groups may experience both inadequate support outside of AI contexts and a sense of dismissal within AI contexts---a potentially disproportional double whammy. 
If bridge conversations are not carefully designed to build preparedness and avoid dependency, reliance on \chatbots for crises could introduce disproportionate exposure to harm.
Governance should therefore require equity audits of crisis use patterns and evaluation of how interventions affect different demographic communities.

\subsection{Limitations and future work}
While our work investigates a critical topic of understanding how people navigate crisis, we understand that this work also has several limitations.
First, our research was limited to a time of crisis only. Our research did not explore other acute states such as delusions or substance abuse contexts. Thus, we do not make recommendation for long term mental health maintenance and care. We understand that longitudinal experiences and complex situations should be studied by crisis experts and may require a different protocol.

Further, we know from prior work in mental health studies that the experiences of disclosing and navigating mental health crisis vary across geographic regions. Our work only considers experiences in the United States where it is not illegal to harm oneself. As AI tools are available around the world, additional studies should consider the lived experience of those across different countries.

We should also highlight that not all survey respondents identified that they were explicitly experiencing suicide. Drawing from on approaches to decolonizing digital mental health~\cite{pendse2022treatment}, we optimized for respondents sharing lived experiences and intentionally chose language in our survey that was colloquial and easy to understand concepts rather than clinical terms (e.g., ``venting'' rather than ``emotional disclosure'').
Given this decision, we designed our interview protocol and data analysis approach with this in mind in order to honor the language of the experiences shared. Our interview participants are also limited to mental health experts and advocates that represent specific subpopulations, not all. Our findings may not be generalizable beyond the specific samples incorporated into our discussion. 

Finally, we highlight that our findings primarily focus on the experience of adults only. Our study included interviews with you advocates, however none of our survey respondents were under 18. Given the unique experiences of young people, future work should extend this study and validate with the lived experiences of adolescents and supplement with the expertise of development psychologists to paint a holistic picture of the experience.

\section{Conclusion}
This research reveals a complex reality that challenges simple narratives about AI in mental health crises. Our participants made calculated decisions shaped by very real barriers in existing mental health infrastructure: stigma, cost, access, and a fear of burdening others. Critically, they were drawn to what they perceived as \nonhuman of conversational AI agents such as infinite patience, non-judgmental listening, immediate availability, and freedom from the emotional labor that human relationships entail. Users are not rejecting human connection, but rather the limitations and costs of seeking help in systems that are already overburdened and under-resourced.

Our adaptation of the stages of change model offers a productive reframing: rather than asking whether AI should intervene in crises, we can ask how AI might support movement toward readiness for meaningful help as a bridge that meets people where they are. But this is only possible if we design with humility about AI's limitations and honestly inquire about the structural failures and opportunities that drive people to seek help from machines.

For the responsible AI community, our findings carry an urgent implication: guidelines cannot be developed in isolation from lived realities of help-seeking. This means moving beyond harm-prevention frameworks that focus solely on what AI should not do, toward affirmative frameworks that articulate genuine support—increasing preparedness for positive action while de-escalating harmful intent. As we develop guidelines for AI agents in mental health crises, let us be mindful to center the voices of those who have lived this experience, design systems that honor their agency, and acknowledging their vulnerability. This will ensure that AI provides the meaningful \textit{bridges}---rather than unthoughtful redirections---that everyone deserves when navigating mental health challenges.

\bibliographystyle{ACM-Reference-Format}
\bibliography{main}

\appendix
\section{Survey Instrument}

\subsection{Eligibility}
\surveyq{Are you at least 18 years old?}
\begin{itemize}
    \item Yes
    \item No
\end{itemize}

\surveyq{Do you have a US-Based mailing address?}
\begin{itemize}
    \item Yes
    \item No
\end{itemize}

\surveyq{Please provide your zip code. This is for survey validation purposes only. Your zip code will be deleted immediately from our records once your response is manually validated by a researcher on our team.}

\surveyq{When was the last time you experienced moment(s) of mental health crisis, characterized as: an acute, overwhelming emotional state that disrupts normal functioning and overwhelms an individual's usual coping mechanisms. In some cases, this may have involved suicidal ideation, self-harming behavior, or severe hopelessness.}
\begin{itemize}
    \item Less than a week ago  
    \item About 1 week ago  
    \item About 2 weeks ago  
    \item About 1 month ago  
    \item About 2-3 months ago  
    \item About 4-6 months ago  
    \item 7 or more months ago  
    \item More than a year ago 
    \item I've never experienced a mental health crisis 
\end{itemize}

\surveyq{Have you used AI conversational agents, such as ChatGPT, for support in a moment of mental health crisis?}
\begin{itemize}
    \item Yes
    \item No
\end{itemize}

\subsection{Demographics} 
Please answer the following questions about yourself and your usage of AI. Identifying information, such as your name and email, is only required for compensation and will be deleted once you have been compensated. 

\surveyq{Please write your name that we can use for your gift card compensation.}

\surveyq{Please write your email that we can use for your gift card compensation.}

\surveyq{What is your age?}
\begin{itemize}
    \item 18 - 25 
    \item 26 - 35 
    \item 36 - 45 
    \item 46 - 55 
    \item 56 - 65 
    \item 66+ 
    \item Prefer not to answer 
\end{itemize}

\surveyq{How do you describe your gender identity? (check all that apply)}
\begin{itemize}
    \item Male 
    \item Female 
    \item Transgender, non-binary, or another gender 
    \item Self described (please specify below)
    \item Prefer not to answer 
\end{itemize}

\surveyq{What is the highest level of education you have attained? }
\begin{itemize}
    \item Less than high school  
    \item High school diploma or equivalent  
    \item Some college or vocational training  
    \item Bachelor’s degree  
    \item Some postgraduate degree or certificate  
    \item Master’s degree  
    \item Doctoral or professional degree  
    \item Prefer not to answer
\end{itemize}

\surveyq{What is your race? Select all that apply and enter additional details in the spaces below.}
\begin{itemize}
    \item American Indian or Alaska Native 
    \item Asian  
    \item Black or African American  
    \item Hispanic or Latino  
    \item Middle Eastern or North African  
    \item Native Hawaiian or Pacific Islander  
    \item White  
    \item Prefer not to answer  
    \item Other 
\end{itemize}

\surveyq{Which category best describes your yearly household income before taxes? Do not give the dollar amount, just give the category. Include all income received from employment, social security, support from children or other family, welfare, Aid to Families with Dependent Children (AFDC), bank interest, retirement accounts, rental property, investments, etc.}
\begin{itemize}
    \item Less than \$5,000  
    \item \$5,000 - \$9,999  
    \item \$10,000 - \$14,999  
    \item \$15,000 - \$19,999  
    \item \$20,000 - \$29,999  
    \item \$30,000 - \$39,999  
    \item \$40,000 - \$49,999  
    \item \$50,000 - \$59,999  
    \item \$60-000 - \$74,999  
    \item \$75,000 - \$99,999  
    \item \$100,000 - \$124,999  
    \item \$125,000 - \$149,999  
    \item \$150,000 or more  
    \item Prefer not to answer 
\end{itemize}

\subsection{General AI Use}
Please answer the following general questions about your experiences with mental health crises and AI

\surveyq{For each of the following resources (e.g., family, friends, etc.) please rate their availability to you in a mental health crisis and the likelihood that you would turn to them. If you have no experience with a resource, please mark "I've never tried."}

Availability Scale:
\begin{itemize}
    \item I've never tried
    \item Never Available
    \item Rarely available
    \item Sometimes Available
    \item Often Available
    \item Always Available 
\end{itemize}

Likelihood Scale:
\begin{itemize}
    \item Almost Never
    \item Rarely
    \item Occasionally (half the time)
    \item Often
    \item Almost Always
\end{itemize}

Resource Options:
\begin{itemize}
    \item Family members 
    \item Friends 
    \item Teachers or coworkers 
    \item Community organizers (e.g., religious leaders, librarians, social workers)
    \item Professional counselors or therapists 
    \item Local and national helplines (e.g., 988, crisis text lines, teen line) 
    \item Social media apps (e.g., TikTok, Instagram, Youtube)
    \item Online peer support forums (e.g., Reddit, Talklife)
    \item Mental health or wellness technologies (e.g., Calm, Headspace)
    \item AI conversational agents (e.g., ChatGPT, Gemini, Character.ai)
\end{itemize}

\surveyq{Is there any resource you use that is not listed above? How available is it to you and what is your likelihood of turning to it in a moment of mental health crisis?}

\surveyq{Which of the following AI conversational agents have you used in a moment of mental health crisis? Please select all that you have tried.}
\begin{itemize}
    \item ChatGPT  
    \item Google Gemini  
    \item Microsoft Copilot  
    \item Perplexity  
    \item Claude  
    \item Deepseek  
    \item Snapchat AI  
    \item Grok  
    \item Replika  
    \item Character.AI  
    \item PI AI  
    \item Woebot  
    \item Wysa  
    \item Youper  
    \item Yana  
    \item Other (please specify)
\end{itemize}

\subsection{Testimony: Your Experience of AI Use for Mental Health Crisis}
In the following set of pages, please tell us your story of turning to AI in a mental health crisis.      For the purposes of this survey, we define a mental health crisis as a serious situation where your mental distress became ovewhelming. This may have posed risks to your safety, well-being, or ability to function. Please only share what you are comfortable with, and use the first-person ``I.''  Please don’t include your name or anyone else’s name in your story. In the following questions, please refer to a most memorable single instance when you turned to an AI agent in a mental health crisis. 

\subsubsection{Tell us your story} Please answer the following questions based on your personal experience.

\surveyq{Describe, in as much detail as you’d like, of a single memorable instance of turning to an AI chatbot in a mental health crisis.}

\surveyq{How long ago did this instance happen?}
\begin{itemize}
    \item Less than a week ago  
    \item About 1 week ago  
    \item About 2 weeks ago  
    \item About 1 month ago  
    \item About 2-3 months ago  
    \item About 4-6 months ago  
    \item 7 or more months ago  
\end{itemize}

\surveyq{Which AI agent did you use in this instance?}
\begin{itemize}
    \item ChatGPT  
    \item Google Gemini  
    \item Microsoft Copilot  
    \item Perplexity  
    \item Claude  
    \item Deepseek  
    \item Snapchat AI  
    \item Grok  
    \item Replika  
    \item Character.AI  
    \item PI AI  
    \item Woebot  
    \item Wysa  
    \item Youper  
    \item Yana  
    \item Other (please specify)
\end{itemize}

\surveyq{What aspects of mental health crisis were you experiencing in this instance? Check all that apply and/or describe it in the text box.}
\begin{itemize}
    \item Suicidal thoughts without any intention to make a plan of action (e.g., wishing you weren't here anymore) 
    \item Suicidal thoughts where you had a plan to end your life 
    \item Thoughts of self-harm (e.g., thinking of what it would feel like to inflict pain on yourself) 
    \item Attempts at self-harm (e.g., cutting, burning, or scratching) 
    \item Inability to function as you typically do, due to mental distress 
    \item Other (please specify)
\end{itemize}

\surveyq{How would you rate your level of satisfaction for the interaction you had with the AI agent during this specific mental health crisis instance?}
\begin{itemize}
    \item Extremely dissatisfied 
    \item Somewhat dissatisfied 
    \item Neither satisfied nor dissatisfied 
    \item Somewhat satisfied 
    \item Extremely satisfied 
\end{itemize}

\subsubsection{Reflective Questions on Your Experience}
Please answer the following reflective prompts based on this specific experience.

\surveyq{Why did you turn to an AI agent for mental health crisis support? What was going on that led you to seeking crisis support from an AI agent?}

\surveyq{Did you consider going to any alternative sources of support, such as family, friends or a mental health provider? Why or why not?}

\surveyq{What did you talk to the AI agent about? What did you say?}

\surveyq{What were you hoping to get out of your interaction with an AI agent? }

\surveyq{Immediately after the interaction with AI, what actions – if any – did you take? For example, did you talk to anyone else? Did you do anything different or special? }

\surveyq{In hindsight, would you have done anything differently?  }

\subsubsection{Reflective Questions on the AI Behavior}
Please answer the following questions based on your experience and opinions about the AI agent's behavior.

\surveyq{Was your interaction with the AI agent helpful? Why or why not?}

\surveyq{Was your interaction with the AI agent hurtful? Why or why not?}

\surveyq{Would you go back to AI agents for the same crisis situation? Why or why not?}

\surveyq{What changes to AI agents would you recommend to better support people in moments of mental health crisis?}

\section{Semi-Structured Interview Guide}
\textit{This guide was a loose scaffold for 60-min interviews with mental health experts. Following best practices for semi-structured interviews, all interviewers were trained to ask relevant follow-up questions and deviate from the guide as needed.}
\subsection{Introduction}
\subsubsection{Researcher Introduction}
Hi, thank you for participating in this interview. My name is [Researcher name] and I’m a [position of researcher]. This project is all about AI uses for managing mental health crises, so I’m going to be asking questions about your experience as a member of [group name] and your personal views about AI and mental health. 
Before we start with the interview questions, I just want to verify your consent in this study

\subsubsection{Consent Check}
I just want to confirm that you received the consent form. To highlight the main points:
\begin{itemize}
    \item Your participation in this study is strictly confidential. The only people who will see this are members of the research team listed on the consent form you signed

    \item At any point in the interview, we can pause, skip a question, or stop altogether. Just let me know what you need
\end{itemize}

\textit{[Note for researcher: Please be on the lookout for any stress, mood or body language of the participant that would necessitate skipping a question, shifting gears, etc. When in doubt, suggest that the participant skip the question.]}

Q: Are there any questions I can answer about the consent form?

\textit{[If consent is confirmed, Researcher can start recording] }

Great! So I’m going to start recording here and then we’ll get into the questions.

\subsection{Participant Context}
We were specifically interested in your perspective given your role as [insert participant role from intake]. 

Q: Can you tell me a bit about your role? Possible Relevant Follow-ups:
\begin{enumerate}
    \item What does advocacy look like in your organization?
    \item Do (you or organization) have a core mission? If so, can you tell me about it?
    \item Do you personally have a philosophy or set of principles about mental health that is important to you?
\end{enumerate}

\subsection{Experiences with Mental Health Crsis \& AI}
We're going to move into our set of questions about mental health crisis and AI. We're intentionally keeping the term ``mental health crisis'' vague, but for AI we're mostly referring to publically available chatbots, like ChatGPT. Because these questions are about a more sensitive topic, I just want to remind you that you're welcome to pause or skip a question at any time.

Q: What does mental health crisis mean to you?

Q: If you're comfortable sharing, what is your experience with mental health crisis? \textit{Note: Possibly clarify either firsthand, secondhand or professional experience.} Possible relevant follow-ups:
\begin{enumerate}
    \item If they mention firsthand experience:
    \begin{enumerate}
        \item Are there resources that you turn to in a mental health crisis?
        \item Have you ever considered turning to AI for support in those moments? Why or why not?
    \end{enumerate}
    \item If they mention secondhand experience (e.g., supporting friends through a mental health crisis) 
    \begin{enumerate}
        \item What were you thinking while navigating that?
    \end{enumerate}
\end{enumerate}

Q: Can you give some examples of helpful responses to someone in a mental health crisis?

Q: Can you give some examples of helpful responses to someone in a mental health crisis?

\subsection{Community Experiences with Mental Health Crisis \& AI}
\textit{This section is mainly geared towards advocates who act as community representatives in an official capacity}

Q: How is mental health crisis unique for [insert community]

Q: What are the general thoughts in your community on using AI to help with mental health crisis?

Q: Are there specific concerns for your community about AI use in mental health crisis?

\subsection{Responsible AI Behavior}
Q: Can you give me an example of a helpful AI response to a user experiencing a mental health crisis?

Q: Can you give me an example of a hurtful AI response to a user experiencing a mental health crisis?
Possible relevant follow-ups
\begin{enumerate}
    \item I noticed you had the AI chatbot speak from an ``I'' perspective, can you say more about that?

    \item I appreciate the focus here on short-term de-escalation. Can you speak to potential mid or long-term impacts?
\end{enumerate}

Q: What do you think a user should know about AI before they turn to an AI chatbot for mental health crisis support?

Q: What are your biggest concerns with people using AI for mental health crisis support?

\subsection{Conclusion}
That’s all the questions I had for you. Is there anything you wanted to talk about that we didn’t cover?
I’m just going to stop the recording and then explain the logistics from here. [stop recording] Thank you again for participating! To confirm, you’ll receive a [dollar amount] gift card from your time today. Is the email you put in our intake form correct?
Additionally, would you like us to email you a list of mental health resources, such as text lines? Are there any final questions I can answer for you before we end?

\end{document}